# On detection and annihilation of spherical virus embedded in a fluid matrix at low and moderate Reynolds number

Ujjawal Krishnam[*,#] and Prafulla K Jha

*Department of Physics, Faculty of Science, The Maharaja Sayajirao University Of Baroda, Vadodara, India 390002*

*Email: ujju-phy@msubaroda.ac.in; pk.jha-phy@msubaroda.ac.in

**ABSTRACT**

Effect of high and low Reynolds number is studied on low frequency vibrational modes of a spherical virus embedded in the aqueous medium. We have used an analytical approach based on fluid dynamic and classical Lamb's theory to calculate the vibrational modes of a virus with material parameters of lysozyme crystal in water. The obvious size effect on the vibrational modes is observed. The estimated damping time which is of the order of picosecond varies with Reynolds number and shows a high value for a critical Reynolds number. The stationary eigenfrequency regions are observed for every quantum number *l* and *n* suggesting the most probable Re ranges for acoustic treatment of viruses in order to detect or annihilate the virus using corresponding virus-water configuration.

# INTRODUCTION

The low frequency vibrational modes of nanometric viruses have been the subject of great concern in last few years[1-9]. Viruses being one of the smallest organisms are the miniscule pockets of proteins containing RNA or DNA surrounded by capsid and have found potential applications in diverse areas of rapidly developing nanotechnologies such as nanotemplates for nanofabrication[10-16] and in diagnosing and treating the viral diseases[17-19]. Besides, nanometric viruses present in aqueous medium can be treated as microstructures which have undoubtly the great fundamental and practical interests driven by potential applications in targeted drug delivery[20]. Nanometric size and approximately spherical shape of many viruses make an analogy with nanocrystals or quantum dots. The analogy between nanometric size and approximately spherical shape of many viruses has been successfully applied to use the highly celebrated Lamb's classical model of confined elastic vibrations in nanomaterials to viruses for the study of their quantized phonon modes[1-9, 17]. The confined elastic vibrations or quantized acoustic waves in nanostructures are mechanical vibrations with frequency of orders of sound divided by a typical length of the nanostructure[2]. Moreover, elastic vibrations of nanoparticles(spherical viruses) manifest themselves in the low frequency Raman Spectra which make determination and understanding of their low frequency vibrational modes possible[17]. The first attempt to find vibrational frequencies of a free, isotropic, homogeneous, continuous sphere using only size, density and speeds of sound in corresponding bulk material was made by Lamb[21]. However, there exist many modifications to this with the inclusion of the effect of surrounding medium[1-3,8-9,22]. A growing interest towards the study of acoustic phonon quantization of viruses is manifested to its suitability in less expensive medical applications [23-25].

The estimation of vibrational frequencies in spherical virus particles or other similar organic nanostructures has been attempted previously by several workers[1-9,21].Most of these studies however have been performed by considering the spherical viruses as bare spherical nanoparticles ignoring the effect of liquid solvent. The effect of surrounding medium on the vibrational modes for any system was first presented by Dubrovskiy[9]. The frequency of the modes turns imaginary in the presence of surrounding medium; while the real part represents the vibration of free particle and the decay of modes due to loss of energy of vibrating particle. With the basic idea of Dubrovskiy, many investigations on the spheres present in the surrounding medium mainly the water or glycerol have been performed[1-3,8-9,22]. These works have treated medium as an elastic medium with zero transverse speed of sound and the viscosity effect was studied for biological objects with an inclusion of torsional modes[26]. This was followed by another study on TMV virus with the consideration of the effect of viscosity on the frequency of vibrational modes[2]. The viscosity of water significantly dampens the free vibrational modes of TMV virus. Though the effect of viscosity is considered successfully, there is no consideration of effects of mutual interaction of inertial and viscous forces on vibrational modes. This can be achieved with an exigent involvement of Reynolds number(Re), a dimensionless quantity which gives the measure of effective inertial forces to viscous forces for given flow conditions, hence describing the nature of flow whether it is laminar or turbulent. Therefore, it is expected that the Reynolds number will aver obvious effects on the vibrational modes of nanometric viruses[27-28]. The convolution of exerted high pressure on virus and its boundary layer condition bolsters the expectation of high values of Reynolds number. Furthermore, the flow around the sphere reaches a critical state characterized by a low drag coefficient and unsteady flow conditions[29]. The principal motion is steady in the case of sufficiently small velocities while with the large Reynolds numbers the laminar flow past a body is unsteady at any rate[30]. Very recently, the role of Reynolds number on the flow of microswimmers or microrobots has been established[31]. Galstyan *et al* [3] have performed a systematic study on the breathing mode of elastic sphere in both Newtonian and Complex media to confirm the experimental observations[1] of viscoelastic trigger of high frequency longitudinal vibrations of bipyramidal nanoparticles. However, their results show contradicting behaviour as far as vibrational mode is concerned. In addition, these works have not fully incorporated with fluid dynamics and vibrational dynamics to understand localised phenomena surrounding fluid embedded virus. However in our previous work[17], we have analogously studied virus as a quantum dot and reported the distinctive involvement of viscoelasticity of fluid matrix in enhancing damping time of virus. We achieved a longer damping time(34.3 ps) for a 50 nm virus in aqueous configuration and predicted an estimation of virus annihilation with principles of rife machine. This striking emergence of damping time as an important parameter provided a direct approach for the detection and killing of the virus particles based on acoustic resonance[3,17]. Damping time as a felicitous parameter in our previous study extrapolates its sheer presence in fluid dynamical consideration of virus-water configuration, and hence to be analytically expressed as a function of Reynolds number. In the present work, we perform an estimate of low frequency vibrational modes of a conformationally stable spherical virus embedded in aqueous medium under the framework of an elastic continuum model entailing the effect of Reynolds number for an improvised understanding of the amenable configuration of virus-surrounding medium for virus destruction. Our calculations show that the presence of surrounding medium water embellishes the damping time to a significantly high value, suggesting a distended likelihood of killing the virus compared with previous estimates.

## METHODOLOGY

The vibrational modes of nanometric spherical virus particle embedded in aqueous medium are calculated using classical Lamb's model[21] and its improved versions[17, 32-38]. Lamb's equation which is actually manifestation of Navier equation in elasticity, gives rise to displacement field.

From Free Shell Model(FSM), the eigenvalue equation for spheroidal mode under the stress free boundary conditions at the surface of the virus are obtained as[32-36]

$$\frac{tan\left(\frac{\omega}{V_l}a\right)}{\left(\frac{\omega}{V_l}a\right)} = \frac{1}{1-\frac{1}{4}\left(\frac{\omega}{V_t}a\right)^2} \text{ for } l=0 \quad (1)$$

$$2\left[\eta^2 + (l-1)(l+2)\left\{\frac{\eta j_{l+1}(\eta)}{j_l(\eta)} - (l+1)\right\}\right]\frac{\xi j_{l+1}(\xi)}{j_l(\xi)} - \frac{\eta^4}{2} + (l-1)(2l+1)\eta^2 + \{\eta^2 - 2l(l-1)(l+2)\}\frac{\eta j_{l+1}(\eta)}{j_l(\eta)} = 0 \text{ for } l \geq 1 \quad (2)$$

Here dimensionless eigenvalue is expressed as

$$\eta_l^S = \frac{\omega_l^S R}{V_t} \text{ and } \xi_l^S = \frac{\omega_l^S R}{V_l} \quad (3)$$

Where ω is the angular frequency and R is radius of an isotropic nanoparticle. $j_l(\eta)$ or $j_l(\xi)$ are the spherical Bessel's function of first kind and $\eta = \xi\left(V_t/V_l\right)$. The torsional mode is a vibration without dilation, and its eigenvalue equation is given by[21]

$$J_{l+1}(\eta) - \frac{l(l-1)}{\eta}J_l(\eta) = 0 \text{ for } l \geq 1(11) \quad (4)$$

The torsional modes are defined for l ≥1 and are orthogonal to the spheroidal modes[27,28]. The eigen frequencies obtained from equations (2) and (3) depend only on single index $l$ which lead to the (2l+1)-fold degeneracies for $l$th vibrational mode. These discrete modes for different values of $l$ are essentially similar to acoustic phonons at discrete **q**-points in the Brillouin zone given by $\pi(l + 1/2)/R$ up to a maximum value π/a at the zone boundary[21]. These modes modify the density of states of bulk[32,33] and have been found responsible for increase in specific heat of nanoparticle at low temperature[39]. The lowest eigenvalue for n=0 for both spheroidal and torsional modes corresponds to the surface modes and have large amplitude near the surface. The frequencies of the spheroidal and torsional modes can be calculated from eigenvalue equations (1,2) and (4) respectively. In order to express the vibrational frequencies in cm$^{-1}$, it is divided by the velocity of light in vacuum c. Therefore, it is possible to observe the modes, allowed by the selection rules in the low frequency Raman scattering. The spheroidal modes are characterised by $l \geq 0$, where $l=0$ is the symmetric breathing mode, $l=1$ is the dipolar mode and $l=2$ is the quadrupole mode. The $l=0$ mode is purely radial and produces polarized spectra, while $l=2$ mode is quadrupolar and produces partially depolarized spectra. The spheroidal modes for even *l (i.e. l=0* and 2) are Raman active[17]. The lowest eigen frequencies for n=0 for both spheroidal and torsional modes correspond to the surface modes while n≥1 corresponds to inner modes as equations are described by quantum number *l* and harmonic *n*.

From Complex Frequency Model(CFM), it is observed that the presence of surrounding medium on the low frequency Raman peaks of spherical viruses is manifested in the line width broadening, shifting of peak positions and even appearance of new peaks. To explain the effect of medium, the equation of motion is solved under new boundary layer conditions and specific acoustic impedance ($Z_m$) at the interface of medium and spherical viruses[17,19]. With an analogy to Complex Frequency Model(CFM), we take into account parameters of spherical virus as it is surrounded by a homogeneous and isotropic matrix of density $\rho_m$ and speeds of sound $V_{lm}$ and $V_{tm}$[17, 26]. The CFM model is the result of fixed boundary condition and is different from the free boundary condition in the sense that it takes into account the stresses at virus boundary, continuity of $\vec{u}$ and force balance at the virus-matrix interface due to the existence of boundaries at the virus particle and matrix interface. The boundaries between the particle and matrix modify the confined phonon modes and even sometimes are responsible for the appearance of new modes. The boundary condition at large R is that $\vec{u}$ is an outgoing travelling wave.

The velocity field of virus particles of radius R is governed by the linear Navier-Stokes equation for compressible flows due to considered small amplitude acoustic waves in the field. Under stress free boundary conditions the Lamb's equation has two kinds of eigenvalues: spheroidal or torsional which are described by orbital angular momentum quantum number *l* and harmonic *n*[17]. We use radius of protein and its average sound velocities to estimate the frequencies of the vibrational modes. On these contours, the presence of surrounding medium water affect the boundary conditions through specific acoustic impedance($Z_m$) and hence modifying eigenvalue equations. The modified eigenvalue equation for the lowest spheroidal mode can be expressed as[19,36]

$$s^2 \tan(s) = \left[ 4 \left( \frac{V_{tp}}{V_{lp}} \right)^2 + is \left( \frac{Z_m}{V_{lp} \rho_p} \right) \right] (\tan(s) - s) \tag{5}$$

Where $V_{tp}$ and $V_{lp}$ are the transverse and longitudinal sound velocities respectively, and $\rho_p$ is the density of virus. $s = \frac{\omega R}{V_{lp}}$ is the eigenvalue and $Z_m$ is the specific acoustic impedance at the matrix interface, which is simply density times the sound velocity for a plane interface. For a spherical interface it is mathematically given by

$$Z_m = \rho_m V_m \frac{\left((k_m R)^2 - i(k_m R)\right)}{1 + (k_m R)^2} \tag{6}$$

Where, $\rho_m$ and $V_m$ are the density and sound velocity in the surrounding medium, respectively. Here $k_m = \frac{\omega}{V_m}$ is the wave vector in surrounding medium. The eigenvalue obtained so can be used to estimate the complex angular frequency ($Im(\omega)$) and then the damping time($\tau_D$) of the normal modes using expression $\tau_D = -1/Im(\omega)$.

This leads us to derive Reynolds number as a function of velocity in the fluid matrix which can be achieved using Free Shell Model, Elastic continuum approach and treatments of fluid dynamics under stress free boundary conditions at the surface of spherical viruses. As investigated by Reynolds[40], Landau[30] and Orszag[41], our approach to fluid dynamics can be extended to relate acoustic phenomena at the surface of virus and Reynolds Number(Re) as a measure of viscous force and inertial force resulting from virus' boundary layer interactions with the surrounding viscoelastic fluid by classical point of view.

Equation for stress by considering a viscoelastic(Kelvin-Voigt) material can be written as[2]

$$\sigma_{ij} = 2(\mu_L + i\omega\mu)\varepsilon_{ij} + (\lambda_L + i\omega\lambda)\nabla \cdot \vec{u}\delta_{ij} \tag{7}$$

Where μ is the shear viscosity, λ is the second viscosity coefficient, $\omega$ is the frequency, $\vec{u}$ is the displacement field, $\varepsilon_{ij}$ and $\sigma_{ij}$ are strain and stress tensors respectively.

And Lamb's equation for three dimensional elastic body in the differential form is given by the expression as,

$$\rho \frac{\partial^2 \vec{D}}{\partial t^2} = (\lambda + \mu)\vec{\nabla}(\vec{\nabla} \cdot \vec{D}) + \mu \nabla^2 \vec{D} \tag{8}$$

Where $\vec{D}$ is a lattice displacement vector, ρ is mass density the two parameters μ and λ are Lame's constants as well as dynamic viscosity coefficient or second viscosity coefficient respectively as Saviot *et al*[5] derives velocity expressions for viscous compressible fluid which stand in the equivalence of velocity expression of Talati *et al* and related to the longitudinal and transverse sound velocities in bulk symbolically as

$$V_t = \sqrt{\frac{\mu}{\rho}} \text{ and } V_l = \sqrt{\frac{2\mu+\lambda}{\rho}} \tag{9}$$

Hence in case of viscous compressible fluid[16],

$$\mu = C_{44} = \frac{Y}{2(1+\sigma)} \tag{10}$$

Where Y and σ are Young's modulus and Poisson's ratio respectively.

As considered in the model, spherical virus is confined and boundary layer is laminar, a fluid may be considered ideal if Re is large and as the rapid decrease in velocity in the boundary layer is ultimately due to the viscosity which counts significantly even if Re is large whereas it is known that very large Reynolds numbers are equivalent to very small viscosities[42]. And with the solutions of Prandtl's equations of motion in laminar boundary layer, we reach at an important result that when Reynolds number is changed, the whole flow pattern in the boundary layer undergoes a similarity transformation, where longitudinal distances and velocities remain unchanged while transverse distances and velocities vary as a function of Re[43].

Now from Reynolds number concept and Orr Sommerfield's stability equation[40-41,44],

We have,

$$Re = \frac{\rho v L}{\mu} \tag{11}$$

Where, ρ is mass density, μ is dynamic viscosity coefficient, L is the distance travelled by fluid and v is maximum velocity an object attains with respect to fluid. As virus traverses through the fluid matrix, velocity of the matrix (water in our study) influences the relative transverse velocity of virus and hence value of *v* in the expression changes relative to the velocity of fluid. Virus has been assumed covering linear distance up to 100 to 1000 times its size as a result of relative fluid movement whereas fluid laminarly covering the distance L in the range of a few nanometers. In case velocity of the fluid is altered, the relative velocity of the object with respect to fluid will also similarly transform, hence shifts will be observed. In that case, combining Eqn (9) and (10),

We have

$$Re = \left(\frac{YL^2\rho}{2\mu^2(1+\sigma)}\right)^{1/2} \qquad (12)$$

As virus is smooth and flow is considered below critical Reynolds number, velocity distribution can be expressed as,

$$v(x,y,z,t) = \sum_{p_1,\ldots,p_n} A_{p_1,\ldots,p_n}(x,y,z)e^{-i\sum_{i=1}^{n}p_i\phi_i} \qquad (13)$$

Where $|A|_{max} \sim \sqrt{Re - Re_{cr}}$ and A is a time function A(t)=const•e$^{-i\Omega t}$, where $\phi_i=\omega_i+\beta_i$ is phase containing n arbitrary initial phases $\beta_i$. As frequencies incommensurate, opting freely a set of simultaneous values for the phases $\phi_i$ will set it to a state beforehand in a long interval of time with the fluid traversing through the states which are apparently close. Character typical of turbulence will develop with certain new sets of periods in succession with further increase of Re. The motion possesses a definite number of degrees of freedom for a definite Re; in the limit as Re tends to infinity, similarly the number of degrees of freedom will grow infinitely large[30]. In subcritical flow, drag coefficient is nearly independent of Reynolds number but an increase in pressure leads to the expectation of high Reynolds number whereas circumferential angle of sphere is also important to be considered when sphere is dragged because of relative fluid motion[29]. Analogy set on the same line with the elastic spherical virus model can improve our understanding of vibrational modes coupled with fluid dynamics understanding involved in the phenomena.

**RESULTS AND DISCUSSION**

Low frequency vibrational modes of nanometric spherical viruses embedded in aqueous medium are investigated with elastic parameters of lysozyme crystal[**Table I**] for an optimum range of sizes and varying Reynolds numbers[17,45-47] by solving eigenvalue equations(1-3) utilising velocities as functions of Re from equation(12) and results are presented in **Table II**. The results assert Reynolds number(Re) as a function of relative translational velocity of virus with respect to water. It can be seen from results that Re significantly modifies eigenvalues and eigenfrequencies. The frequencies of torsional modes remain unchanged with varying Reynolds number, indicating their purely transverse nature and independence of torsional modes from material property. Eigenfrequencies corresponding to *l*=0 and *n*=0 spheroidal mode are presented in last column of **Table II,** and it is obvious that values are varying with size of virus for a given Reynolds number. The energy eigenvalues for virus particles of different sizes show blueshift of spheroidal mode energy with decreasing size which is quite consistent with our previous study[17] and other studies[1-9]. As the effect of water on biological structure has been earlier reported[48-50], our result shows significant variation in energy eigenvalues with the involvement of viscous and inertial forces and this is majorly because of role of water as a Newtonian fluid. The effect of water is to dampen the modes and turning the frequency into the complex frequency. This implies the attenuation of oscillation of virus in the presence of surrounding medium[3] which is clearly visible in **Fig. 1** and **Fig. 2**. This attenuation in oscillation results from propagation of energy of modes into the surrounding medium away from vibrating virus particles through travelling waves [51]. It is quite fascinating to observe the stationary regions for every $l$ and $n$ in **Fig. 1**, these regions suggest criticality of the flow around the virus and the best probable scenario to annihilate the virus with acoustic treatment.

For an undamped oscillation system, damping time($\tau_D$) is infinite, so is the quality factor(Q) but for a system experiencing damping, condition slightly differs. In case system suffers small damping, resonant frequency is nearly equal to natural frequency $\omega/2\pi$. In a fluid system, virus being in oscillation is hindered by water surrounding it where water acts as an external agent to alter the vibrational frequency of virus and damping time. **Fig. 2** illustrates the suppressing of eigenfrequencies with an increase in Re for varying virus sizes whereas stationary regime is seen to be more significant with a conglomerated virus size. This can be well explained with the inclusion of the concept of relative damping. We observe that as Reynolds number increases, the increase in relative viscous action results into shifting of velocity opposing viscous forces, hence damping force comes playing a major role for virus-water configuration. The measure of water's velocity and relative mass measure of virus determine the magnitude of damping. As the mass of virus particle increases, damping effect($\zeta = c/2\sqrt{km}$) is relatively altered. For a virus of smaller size and mass, relative movement of water with respect to virus endows a greater damping. As Re manifests within the limits of critical Reynolds number($Re_{cr}$), relaxation time appears to be slowly elevating the time virus-water system elapses to decay to $e^{-1}$ value. This process continues until system achieves almost static configuration. In case the relative movement around virus particle exceeds critical range, the system plunges to suffer very significant damping and equilibrium is largely hampered which is observed to be resulting into steep decrease in damping time with further increase in Reynolds number beyond $Re_{cr}$. This finally resorts to our results that the increase in Reynolds number as a relative measure of viscous force to inertial force, results into an effective suppression of eigenfrequencies in case of all viruses of varying radii. With $Re \approx Re_{cr}$, the nonstationary motion is difficult to materialize; and in case poiseuille motion with Re below $Re_{cr}$ is adopted, motion becomes somewhat steady with respect to significantly large perturbations for varying modes even in the case $Re_{cr} > Re$. In subcritical flow, drag coefficient remains nearly independent of Re, so eigenvalues are not sufficiently affected in region of Re≈Re$_{cr}$ and is independent of virus sizes as variation of eigenfrequencies appear significant. One of the major aims of present study is to find the critical Reynolds number(Re$_{cr}$) for which damping time is significantly high so that combination of virus and water can be used to detect and kill viruses and this is quite evident from stationary regions of **Fig. 2**. It is interesting to note that the damping time smoothly increases for the value of Re and becomes maximum for Re≈Re$_{cr}$ and further rapidly comes to zero. Damping time remains zero after critical Reynolds number Re$_{cr}$. As studied by Djellouli *et al*[31], it is found that shape hysteresis results in quasisimilar stability in shape deformations with displacement of 1% of radius of spherical microrobot( a self propeller unlike virus which swims as a result of relative fluid disturbance). This enhances the stability for acoustic treatment on viruses, even if they become conformationally instable. This supports resonance phenomenon to a favourable peak and incorporates with inertial interaction between fluid and virus even at small scales.     Keeping in view that the fluid movement past virus no matter how small is necessary, the damping time of virus with varying radii has been estimated and it is observed that damping time remains same for all velocity values within the limits of acoustic V$_t$ and V$_l$ values of virus, and results are presented in **Table III**. On the other hand, damping time drastically increases with an increase in relative velocity values beyond acoustic velocities of virus. For an enhanced picture of this phenomenon, we selectively study the variation of damping time with Reynolds Number for the virus of radius 56nm and calculated data from equation (5-6) are presented graphically in **Fig. 3**. Here, damping time is observed to be increasing with size; however it remains in the range of picoseconds. This is consistent with the protein-solvent coupling time scale occurring in both relaxations and low frequency vibrations of proteins[52]. A comparison has been drawn for spheroidal eigenvalues without inclusion of Reynolds number, i.e. without using equation (12) and just with standard parameters from **Table I**, against ones expressed as functions of Re, and presented in **Figure 4**. Results show that with an increase in Reynolds number, spheroidal modes for harmonic *n* appear significantly regressive. Stronger hydrogen bonds at viral protein surface could explain the reason behind regression [53]. The study supports extension for linear chains of spherical viruses as propelling sphere cargo at low Re[4].

**CONCLUSION**

The present paper reports the Reynolds number dependency of low frequency vibrational modes and damping time of a spherical virus in reference of virus detection and destruction. Study shows that for a virus of fixed radius, damping time does not change for a system unless induced relative velocity exceeds the parametric velocity of virus. As high acoustic mismatch in virus-matrix configuration is observed, the virus can be destroyed by considering the role of Re as a function of viscous and inertial forces evident in viscoelastic virus embedded system where mechanical annihilation of the virus can be achieved with the principle of rife therapy where resonance between microbe's lethal mechanical oscillation frequency and mechanical oscillation frequency would facilitate the process.

**CONFLICT OF INTEREST**

There is no conflict of interest.

**ACKNOWLEDGEMENT**

**TABLE CAPTION:**

**Table I:** The acoustic parameters for spherical virus embedded in aqueous medium.

**Table II:** Eigenmodes and Eigenfrequencies of Spherical Virus with varying Reynolds number.

**Table III:** Damping time of viruses in aqueous medium for vibrations pertaining to virus-water. configuration within the range of standard acoustic parameters.

**FIGURE CAPTION:**

**Fig. 1:** Effect of Reynolds number on Spheroidal Eigenvalues.

**Fig. 2:** Effect of Reynolds number on Eigenfrequencies of spherical viruses of varying radii.

**Fig. 3:** Effect of Reynolds Number on damping time of spherical virus of radius 56nm.

**Fig. 4:** Spheroidal mode eigenvalues with or without Reynolds number consideration for **(a)**$l$=0 **(b)** $l$=1**(c)**$l$=2 for varying *n.*

Table: I

|  | $V_l$(m/s) | $V_t$(m/s) | ρ(m/s) | $\frac{\rho_m}{\rho_{Virus}}$ |
|---|---|---|---|---|
| Water | 1483 | 0 | 1.00 | 0.826 |
| Virus | 1817 | 915 | 1.21 | 1 |

Table II:

| | | | | | | | Eigenfrequencies(cm⁻¹) of a spherical virus | | | |
|---|---|---|---|---|---|---|---|---|---|---|
| $v_t$(m/s) (Transverse velocity of virus) | | Spheroidal modes | | | Torsoinal Modes | | | Eigenfrequency(cm⁻¹) Of Virus with Radius | | | |
| | | l | n | Eigenvalue with Reynolds consideration | l | n | Eigenvalue with Reynolds consideration | 5628nm | 562.8nm | 56.28nm | 5.628nm |
| 700 | Re=1536.04138 | 0 | 0 | 0.792000 | 1 | 0 | 5.764000 | 0.005 | 0.047 | 0.468 | 4.68 |
| | | | 1 | 6.779000 | | 1 | 9.096000 | | | | |
| | | | 2 | 10.680000 | | 2 | 12.323000 | | | | |
| | | 1 | 0 | 0.874000 | 2 | 0 | 2.502000 | | | | |
| | | | 1 | 4.814000 | | 1 | 7.137000 | | | | |
| | | | 2 | 7.359000 | | 2 | 10.515000 | | | | |
| | | 2 | 0 | 1.717000 | 3 | 0 | 3.865000 | | | | |
| | | | 1 | 2.907000 | | 1 | 8.445000 | | | | |
| | | | 2 | 6.517000 | | 2 | 11.882000 | | | | |
| 800 | Re=1751.35681 | 0 | 0 | 0.618000 | 1 | 0 | 5.764000 | 0.004 | 0.044 | 0.442 | 4.423 |
| | | | 1 | 6.731000 | | 1 | 9.096000 | | | | |
| | | | 2 | 10.610000 | | 2 | 12.323000 | | | | |
| | | 1 | 0 | 0.683000 | 2 | 0 | 2.502000 | | | | |
| | | | 1 | 4.798000 | | 1 | 7.137000 | | | | |
| | | | 2 | 7.351000 | | 2 | 10.515000 | | | | |
| | | 2 | 0 | 1.602000 | 3 | 0 | 3.865000 | | | | |
| | | | 1 | 2.890000 | | 1 | 8.445000 | | | | |
| | | | 2 | 6.498000 | | 2 | 11.882000 | | | | |
| 900 | Re=1967.55872 | 0 | 0 | 0.605000 | 1 | 0 | 5.764000 | 0.004 | 0.041 | 0.406 | 4.064 |
| | | | 1 | 6.728000 | | 1 | 9.096000 | | | | |
| | | | 2 | 10.606000 | | 2 | 12.323000 | | | | |
| | | 1 | 0 | 0.670000 | 2 | 0 | 2.502000 | | | | |
| | | | 1 | 4.797000 | | 1 | 7.137000 | | | | |
| | | | 2 | 7.351000 | | 2 | 10.515000 | | | | |
| | | 2 | 0 | 1.594000 | 3 | 0 | 3.865000 | | | | |
| | | | 1 | 2.889000 | | 1 | 8.445000 | | | | |
| | | | 2 | 6.497000 | | 2 | 11.882000 | | | | |
| 1000 | Re=2191.83203 | 0 | 0 | 0.540000 | 1 | 0 | 5.764000 | 0.003 | 0.035 | 0.35 | 3.499 |
| | | | 1 | 6.713000 | | 1 | 9.096000 | | | | |
| | | | 2 | 10.585000 | | 2 | 12.323000 | | | | |
| | | 1 | 0 | 0.598000 | 2 | 0 | 2.502000 | | | | |
| | | | 1 | 4.792000 | | 1 | 7.137000 | | | | |
| | | | 2 | 7.348000 | | 2 | 10.515000 | | | | |
| | | 2 | 0 | 1.556000 | 3 | 0 | 3.865000 | | | | |
| | | | 1 | 2.884000 | | 1 | 8.445000 | | | | |
| | | | 2 | 6.492000 | | 2 | 11.882000 | | | | |
| 1100 | Re=2403.9244 | 0 | 0 | 0.490000 | 1 | 0 | 5.764000 | 0.003 | 0.027 | 0.265 | 2.652 |
| | | | 1 | 6.703000 | | 1 | 9.096000 | | | | |
| | | | 2 | 10.570000 | | 2 | 12.323000 | | | | |
| | | 1 | 0 | 0.543000 | 2 | 0 | 2.502000 | | | | |
| | | | 1 | 4.789000 | | 1 | 7.137000 | | | | |
| | | | 2 | 7.346000 | | 2 | 10.515000 | | | | |
| | | 2 | 0 | 1.529000 | 3 | 0 | 3.865000 | | | | |
| | | | 1 | 2.881000 | | 1 | 8.445000 | | | | |
| | | | 2 | 6.488000 | | 2 | 11.882000 | | | | |

Table III

| Radius=5.628nm | Radius=56.28nm | Radius=562.8nm | 5628nm |
|---|---|---|---|
| 3.97ps | 30.97ps | 309.7ps | 3097ps |

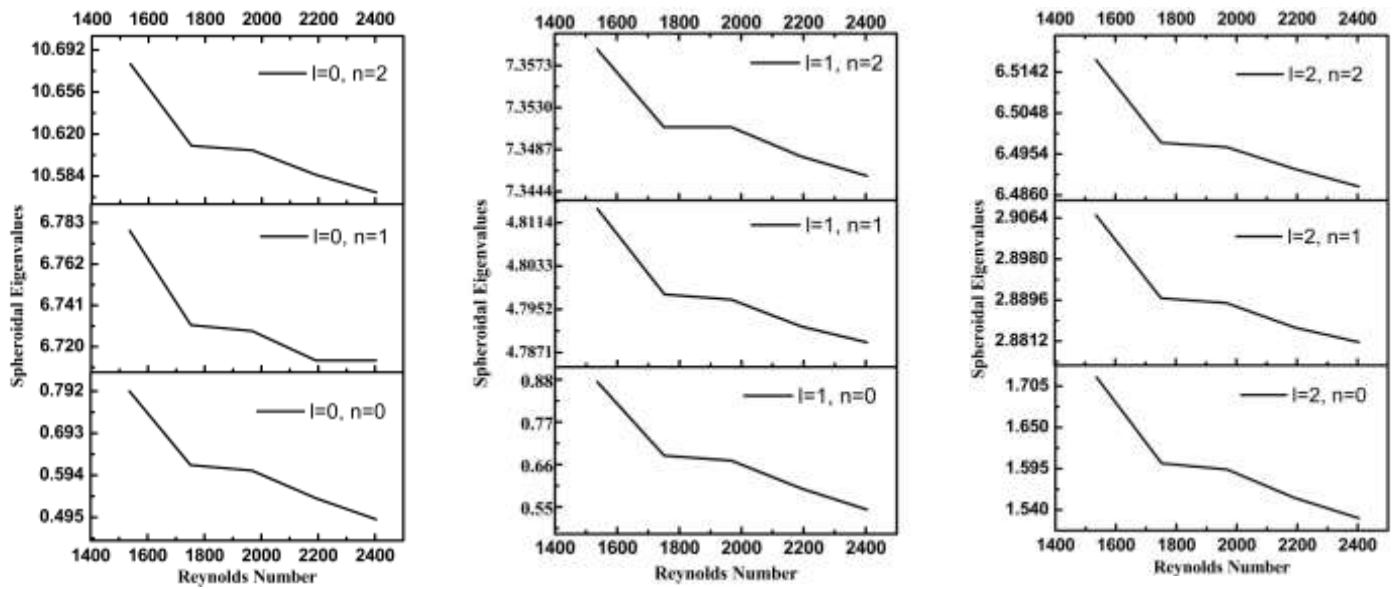

**Fig. 1:** *Krishnam et al*

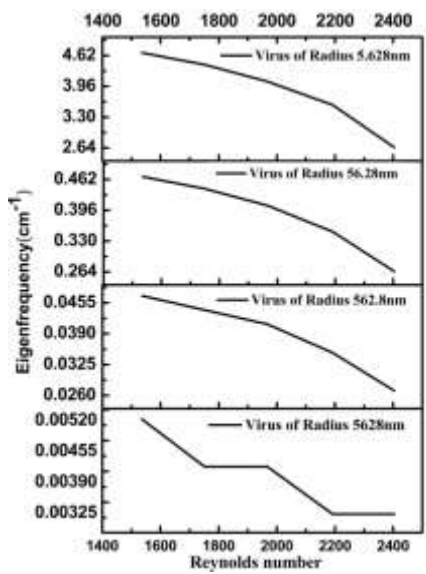

**Fig. 2:** *Krishnam et al*

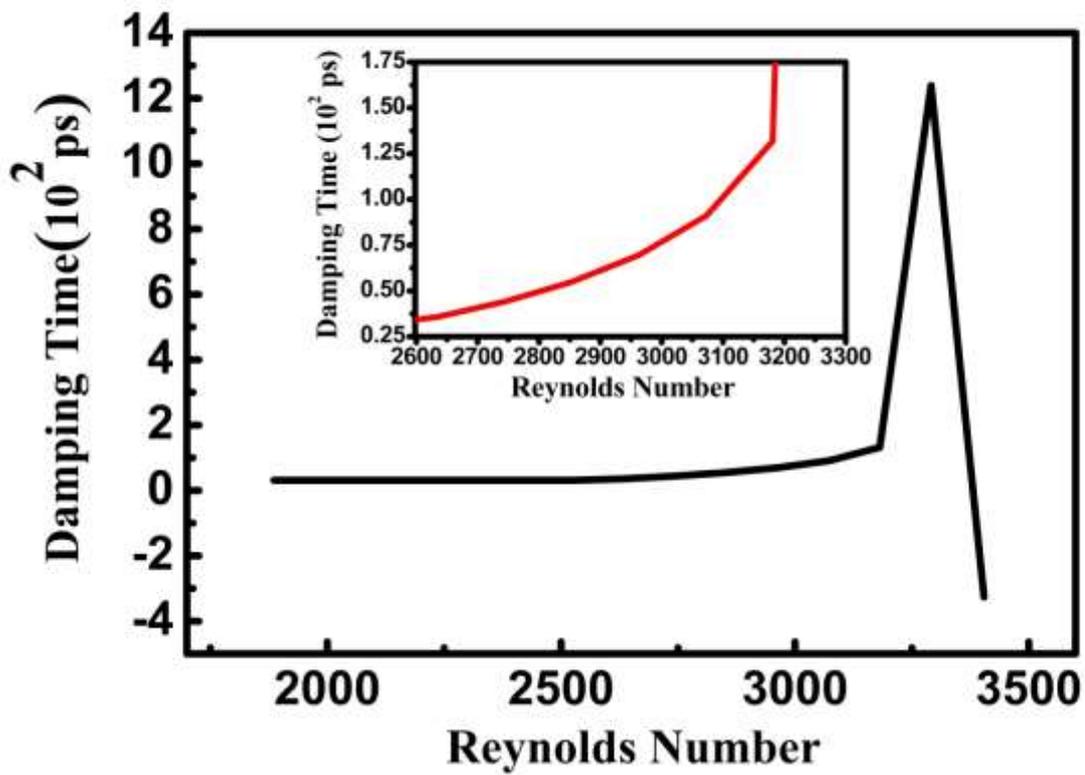

**Fig. 3**: *Krishnam et al*

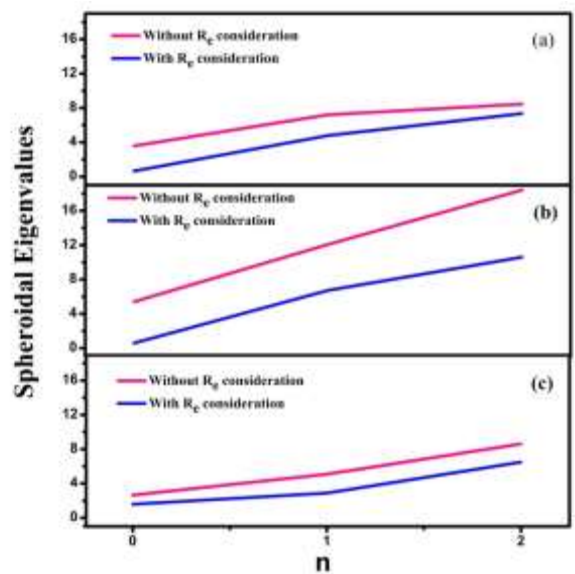

**Fig. 4**: *Krishnam et al*